\documentclass[conference]{IEEEtran}
\usepackage{graphicx}%------------------------------added by ZY
\usepackage{subfigure}
\usepackage{float}%---------------------------------added by ZY
\usepackage{mathrsfs}%------------------------------added by ZY
\usepackage{amsfonts}
\usepackage{array}
\usepackage{amssymb,amsmath}
\usepackage{longtable}
\usepackage{rotating}
\usepackage{multirow}
\usepackage{epstopdf}
\usepackage[lined,boxed,commentsnumbered, ruled]{algorithm2e}
\usepackage{cite}
\usepackage{float}
\usepackage{stfloats}
\usepackage[colorlinks,
            linkcolor=black,
            anchorcolor=black,
            urlcolor=blue,
            citecolor=black]{hyperref}
\usepackage[hyphenbreaks]{breakurl}

\renewcommand{\baselinestretch}{1}
\def\BibTeX{{\rm B\kern-.05em{\sc i\kern-.025em b}\kern-.08em
    T\kern-.1667em\lower.7ex\hbox{E}\kern-.125emX}}

\renewcommand{\baselinestretch}{1}
\begin{document}

\title{Federated Transfer Learning Aided Interference Classification in GNSS Signals}
\author{\IEEEauthorblockN{Min Jiang\textsuperscript{1}, Ziqiang Ye\textsuperscript{1}, Yue Xiao\textsuperscript{1} and Xiaogang Gou\textsuperscript{2}}
\IEEEauthorblockA{\textsuperscript{1}{National Key Laboratory of Wireless Communications} \\
{University of Electronic Science and Technology of China, Chengdu, 611731, China}
\IEEEauthorblockA{\textsuperscript{2}{The 54th Research Institute of China Electronics Technology Group Corporation, Shijiazhuang, 050050, China.}}
{Email: xiaoyue@uestc.edu.cn}
}
}

\maketitle

\begin{abstract}
This study delves into the classification of interference signals to global navigation satellite systems (GNSS) stemming from mobile jammers such as unmanned aerial vehicles (UAVs) across diverse wireless communication zones, employing federated learning (FL) and transfer learning (TL).
Specifically, we employ a neural network classifier, enhanced with FL to decentralize data processing and TL to hasten the training process, aiming to improve interference classification accuracy while preserving data privacy.
Our evaluations span multiple data scenarios, incorporating both independent and identically distributed (IID) and non-identically distributed (non-IID), to gauge the performance of our approach under different interference conditions. Our results indicate an improvement of approximately $8\%$ in classification accuracy compared to basic convolutional neural network (CNN) model, accompanied by expedited convergence in networks utilizing pre-trained models.
Additionally, the implementation of FL not only developed privacy but also matched the robustness of centralized learning methods, particularly under IID scenarios.
Moreover, the federated averaging (FedAvg) algorithm effectively manages regional interference variability, thereby enhancing the regional communication performance indicator, $C/N_0$, by roughly $5\text{dB}\cdot \text{Hz}$ compared to isolated setups.
% Utilizing a convolutional neural network model, specifically the VGGnet acquired through transfer learning, our experiments achieved a notable classification accuracy of $96.69\%$.
% Additionally, this approach effectively preserves the communication performance metric, $C/N_0$, by approximately $6\text{dB}\cdot \text{Hz}$ over the solo scenario.
% The findings suggest that the classification accuracy under the federated learning framework is closely aligned with that of centralized learning models, despite the non- independent and identically distributed nature of client datasets and the presence of interference, underscoring the potential of federated learning in managing GNSS interference signals while ensuring user privacy and maintaining communication performance.
\end{abstract}


\section{Introduction}
% briefly introduce the importance of jamming and anti-jamming
% how to recognize the jammer in conventional way and what is the flaw (privacy concern, efficiency, etc.)
% introduce federated learning
% explain why federated learning will be used in current research
% highlight the contribution of our works

%介绍干扰对GNSS系统的影响
GNSS receivers are widely used in both industrial and civilian domains, thanks to their all-weather capability, robust real-time performance, and precise navigation and timing\cite{silva2023comprehensive}.
However, the GNSS jamming signals, particularly within the L-band spectrum, can disrupt a GNSS receiver to the point of disabling its operation\cite{morales2019jammer}.
%The availability of a broad range of inexpensive jammers on the market increases the risk of deliberate human-made jamming \cite{elghamrawy2021high}.
Research has identified jamming as the primary reason for outages in GNSS-based services\cite{gao2016protecting}.
Thus, incorporating defenses against such attacks is considered an essential feature for GNSS receivers\cite{wu2023jammer}.
Furthermore, the considerable distance of GNSS satellites from Earth reduces the signal power reaching the receiver due to path loss, rendering it highly susceptible to interference\cite{grant2009gps}, while most intentional interferences originate from personal privacy devices, which, despite being illegal, are readily available online\cite{elghamrawy2021high}.
Therefore, these interference devices may cause incorrect positioning, navigation and timing services. As results, interference signals can lower receiver performance by decreasing the carrier to noise ratio ($C/N_0$), potentially leading to service interruptions.

%Thus,the presence of interference signals can easily disrupt GNSS receivers' normal operations.
%According to \cite{silva2023comprehensive}, interference can be divided into natural, caused by satellite, atmospheric and receiver errors, and human-made.
%Human-made interference can be further divided into intentional (such as jamming, spoofing and eavesdropping) and unintentional (caused by RF system harmonics, electrical equipment emissions, and signal leakage).

%随后介绍减轻干扰的步骤，目前对于干扰消除的一些研究，干扰分类对于干扰消除的重要性
With the increasing prevalence of interference signals, safeguarding GNSS receivers against such disruptions has become imperative. 
% Interference mitigation typically involves three steps: detection, characterization, and mitigation \cite{silva2023comprehensive}.
% Recognizing interference is paramount for effective mitigation. 
Accurate interference recognition enables the detection, characterization, and subsequent elimination of interference, utilizing interference parameter models in tandem with relevant techniques.
%Detection techniques encompass both pre-correlation and post-correlation methods \cite{silva2023comprehensive}.
%Anti-interference technologies are categorized into front-end, pre-correlation, and navigation-level approaches, depending on where the mitigation techniques are applied within the receiver \cite{silva2023comprehensive}. 
Recently, solutions for interference mitigation have expanded to detection\cite{axell2015jamming}, mitigation\cite{mao2016robust}, localization\cite{strizic2018crowdsourcing}.
In \cite{mosavi2016narrowband}, the authors proposed a predictor based on neural networks for global positioning system (GPS) anti-interference.
Additionally, \cite{arjoune2020novel} assessed machine learning algorithms, including support vector machine (SVM), neural network, and random forest (RF), for detecting interference in wireless communication systems. Furthermore, authors in \cite{wei2023radar} proposed an intelligent radar anti-interference decision-making method based on deep deterministic policy gradient and multi-agent deep deterministic policy gradient to cope with changes in the interference aircraft's strategies.

While considerable progress has been made in interference mitigation, research specifically targeting interference classification remains relatively sparse. 
%In \cite{mosavi2016narrowband}, the authors proposed a predictor based on neural networks for GPS anti-interference.
%\cite{arjoune2020novel} assessed machine learning algorithms, including support vector machine (SVM), neural network, and random forest (RF), for detecting interference in wireless communication systems.%without involving classification.
%Authors in \cite{wei2023radar} proposes an intelligent radar anti-interference decision-making method based on deep deterministic policy gradient and multi-agent deep deterministic policy gradient to cope with changes in the interference aircraft's strategies.
%目前一些干扰分类方法，以及他们在收集数据时的弊端，引入联邦学习。
In \cite{morales2019jammer}, the authors used SVM and CNN for interference classification, both methods achieved accuracies over $90\%$.
Building upon this work, \cite{swinney2021gnss} enhances GNSS interference signal classification accuracy using signal representation concatenation and transfer learning, evaluating classifiers like SVM, RF, and logistic regression (LR).
%Furthermore, \cite{wu2023jammer} utilizes federated learning (FL) for classifying GNSS interference signals, comparing it to centralized learning.
Most GNSS interference research relies on synthetic data, however, collecting authentic data is crucial for training effective GNSS interference classifiers. Traditional crowdsourcing methods, which often require clients to record and directly share data with central servers, pose significant privacy threats\cite{wu2023jammer}.
Federated learning (FL) addresses this challenge by decentralizing data, enabling models to be trained on local devices and transmitting only updated model parameters to a central server.
Thus, leveraging FL in the classification network enables indirect data exchange between clients and the central server, significantly enhancing both resource efficiency and privacy.

%一些在干扰分类领域联邦分布式学习研究,本文研究
%The authors in \cite{nicola2020collaborative} demonstrated the effectiveness of collaborative management in handling GNSS interference.
The effectiveness of collaborative management in mitigating GNSS interference was demonstrated by the authors in \cite{nicola2020collaborative}.
%Besides, authors in \cite{wu2023jammer} achieved similar results in interference classification through federated learning, highlighting its comparability to centralized learning.
%Furthermore, the authors in \cite{wu2023jammer} employed FL for interference classification, achieving performance comparable to that of centralized learning.
Furthermore, FL was utilized for interference classification in \cite{wu2023jammer}, achieving performance comparable to centralized learning.
However, none of these studies dealt with the identified interference or analyzed the impact of interference on the performance of the communication system. For addressing the above-mentioned issue,the contributions of this paper are summarized as follows:
\begin{itemize}
\item We propose a neural network-based interference classifier that prioritizes data privacy via federated learning (FL) and accelerates training using transfer learning (TL).
\item Within a wireless communication area, 
%each local client, using only a small set of training data, effectively handles changes in interference strategies and completes classification tasks by exchanging model parameters with the central server, thereby minimizing the impact on the GNSS receiver performance metric $C/N_0$.
each local client efficiently manages changes in interference strategies and accomplishes classification tasks using a small training dataset. This is achieved by exchanging model parameters with the central server, thereby minimizing the impact on the GNSS receiver performance metric $C/N_0$.
\item We evaluate the effectiveness of our interference classifiers under various data distribution scenarios, including both independent and identically distributed (IID) and non-IID settings.
\end{itemize}

The paper is structured as follows: Section II outlines the system model for interference classification in GNSS and provides an overview of different types of jamming. Section III elaborates on the FL technique utilized. Simulation results are presented in Section IV, followed by conclusions drawn in Section V.

%Besides,we aim to enable each local client to effectively perform interference classification tasks, even with limited interference signal categories in their training dataset.This approach ensures resilience against changes in interference types and seeks to minimize the impact of interference signals on GNSS receiver performance metrics, such as $C/N_0$.Utilizing FL, clients only need to transmit model parameters, not exchange raw data directly with the central server.This mechanism ensures data privacy and enhances security during the data collection process.Furthermore, by exchanging model parameters, a client encountering a new type of interference can use knowledge from others’ model parameters to successfully classify it.As in \cite{wu2023jammer}, we evaluated the effectiveness of interference classifiers across various data distribution scenarios, including both independent and identically distributed (IID) and non-IID settings.

\begin{figure*}[h]
\centering
\includegraphics[width=1\textwidth]{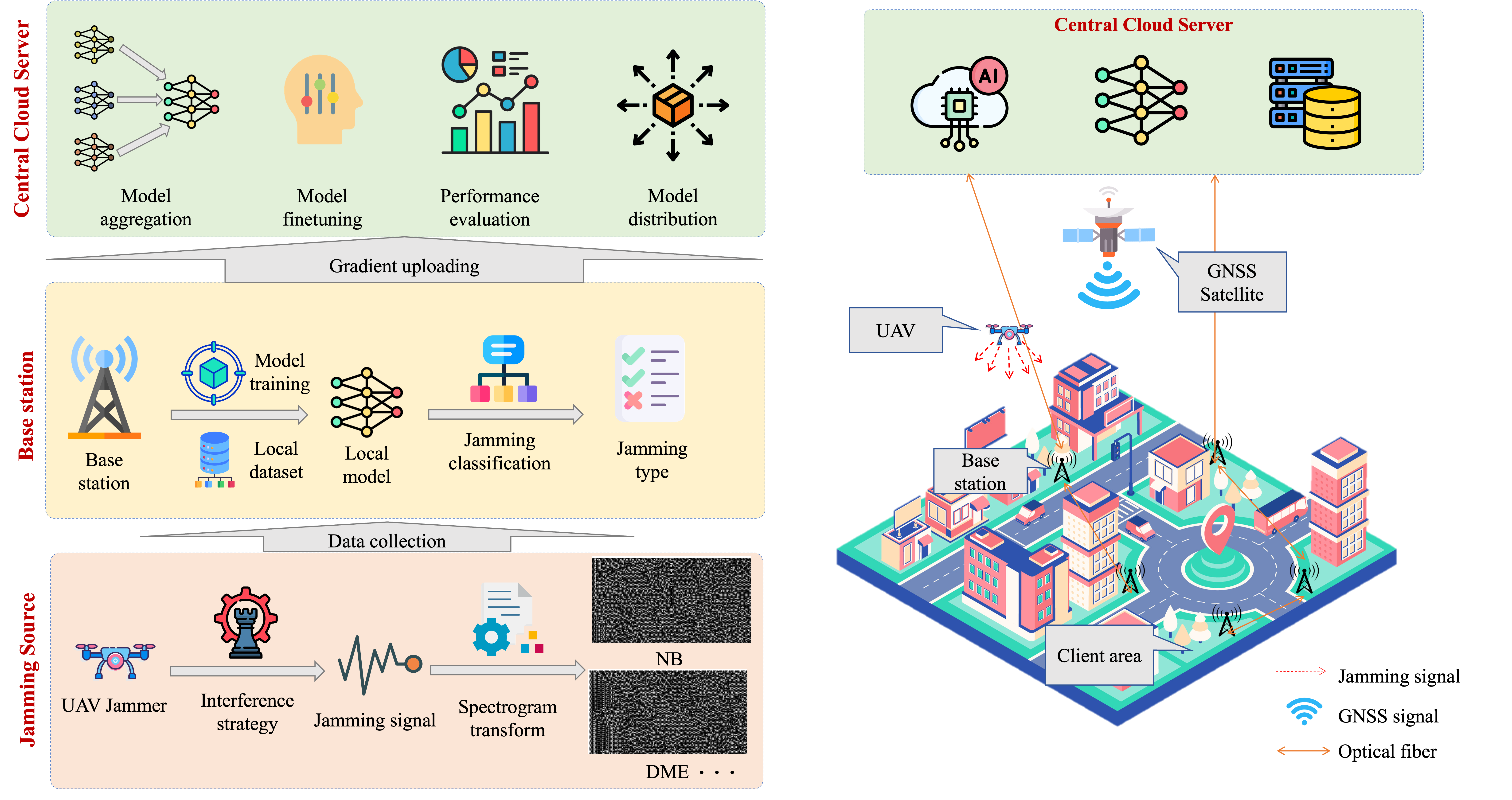}
\caption{A simple system model for jamming signal classification: A movable jammer, such as unmanned aerial vehicle (UAV), generates interference signals across regions, employing diverse interference strategies.
}
\label{fig:1}
\end{figure*}
\section{System Model}  
\subsection{System Outline}
In this article, we focus on the potential occurrence of signal interference in $M$ wireless communication zones, denoted as $\mathcal{M}=\{1,2,\cdots,M\}$. Each zone is connected to a central cloud server (CCS), which facilitates information sharing with each base station  across  each zone.
As shown in Fig. \ref{fig:1}, we utilize a UAV as a mobile jammer, serving as the source of interference. This UAV potentially disrupts wireless communication signals in these zones through various strategies denoted as $\mathcal{L}$.
% In this article, our focus is solely on the various categories of interference.
The interference strategy $l_m \in \mathcal{L}$ varies with different zone $m$, including different types of signals transmitted within the zones.
This means that the jamming UAV may emit a specific or multiple jamming signal types in one specific area, while in another area, the UAV may transmit a completely different signal.
We assume the UAV's position varies within each zone, generating interference $j_m$ in zone $m$, where $j_m \in \mathcal{J}$, which varies due to different interference strategies $l_m$.    
Each zone is equipped with interference detectors constructed with convolutional neural networks (CNNs), denoted as $D_m$, where $m$ represents the zone index.
Employing the FL strategy, base station $B_m$ in zone $m$ shares network model parameters with the CCS to accomplish the task of interference recognition.
Subsequently, based on the recognition results, appropriate measures are implemented to mitigate the interference.

The presence of interference signals amplifies the system's cumulative noise power, thereby reducing the $C/N_0$ and adversely affecting communication system performance\cite{poisel2011modern}.
Precise characterization of interference types enables a neural network classifier to initiate appropriate countermeasures against interference signals, mitigating their adverse effects on the $C/N_0$ of communication systems.
%counteracting the negative impact on the communication systems' $C/N_0$.

If the jammer is not detected under a specific region $m\in \mathcal{M}$, we define the carrier-to-noise ratio under interference $j_m$ as $C/{N_0}_{m,j}$.
Otherwise, $\tilde{C/{N_0}_{m,j}}$ in the absence of interference, which includes conditions following successful interference recognition and elimination.
Consequently, the corresponding $C/{N_0}_{m}$ is calculated as follows:
\begin{equation}\label{eq:1}
    C/{N_0}_{m} = \sum_{j_m\in \mathcal{J}}\eta \tilde{C/{N_0}_{m,j}} + (1-\eta) C/{N_0}_{m,j},
\end{equation}
where $\eta$ denotes the classification recognition accuracy.

Following \cite{ding2023performance}, the relationship between $C/{N_0}_{m,j}$ and $\text{SNR}$ is calculated as follows:
\begin{equation}\label{eq:2}
    C/{N_0}_{m,j} = \text{SNR}_{m,j} \times B,
\end{equation}
where $\text{SNR}_{m,j}$ represents signal-to-noise ratio for base station $B_m$ under undetected interference $j_m$, $B$ denotes the bandwidth of the received signal for $B_m$.
Similarly, we can calculate the $\tilde{C/{N_0}_{m,j}}$ as
\begin{equation}\label{eq:2}
    \tilde{C/{N_0}_{m,j}} = \tilde{\text{SNR}_{m,j}} \times B,
\end{equation}
where $\tilde{\text{SNR}_{m,j}}$ represents signal-to-noise ratio for $B_m$ under detected interference $j_m$.
% However, due to the influence of interference signals, the effective $C/N_0$ is used to describe the signal quality in the presence of interference.
% The equation is shown as follows:
% \begin{equation}\label{eq:2}
%     (C/{N_0})_{eff} = \frac{1}{\frac{N_0}{P_s}+\frac{P_J/P_s}{Qr_c}},
% \end{equation}
% where $r_c$ represents the pseudo-code rate , the variable $Q$ is related to the filter, GPS spectrum, and interference spectrum, and $P_s/N_0$ and $P_J/P_s$ are the traditional $C/N_0$ and $J/S$.

\subsection{Classification of Jamming}
We roughly model the received signal as follows: 
\begin{equation}\label{eq:2}
    r(t)=s(t)+j(t)+w(t).
\end{equation}
where $s(t)$ represents the useful signal,  
$j(t)$ denotes the interference signal generated by UAV interference sources, and we assume the noise signal is additive white Gaussian noise (AWGN), represented by $w(t)$.
If there is no interference, $j(t)$ is set to $0$.
This paper utilizes the interferences provided in \cite{morales2019jammer}, which include various signal types:

{\bf\em 1) Amplitude modulation (AM) jamming: }
it is expressed as:
\begin{equation}\label{eq:3}
    j(t) = \sum_{k=1}^{n} \sqrt{P_{J_k}} e^{j(2\pi f_{J_k} t + \theta_{J_k})}.
\end{equation}
AM interference manifests as continuous wave (CW) interference. 
When $n=1$, it constitutes single-tone interference, while for $n>1$, it transforms into multi-tone interference.
Here, $P_{J_k}$ represents the power of the $k^{th}$ interference component, $f_{J_k}$ signifies its corresponding frequency and $\theta_{J_k}$ is the phase.

{\bf\em 2) Chirp jamming: }
it is expressed as:
\begin{equation}\label{eq:4}
    j(t) = \sqrt{P_J} e^{j\left(2\pi f_J t + \pi b \frac{(f_{\text{max}} - f_{\text{min}})}{T_{\text{swp}}}{t^2} + \theta_J\right)}.
\end{equation}
This is a type of signal whose frequency is linearly modulated with time, achieved by scanning frequencies over a certain time range and frequency range. $b=1$, it indicates upward linear frequency modulation, whereas $b=-1$ signifies downward linear frequency modulation.
$P_J$ denotes the interference power, $f_J$ represents the interference frequency, $f_{\text{min}}$ and $f_{\text{max}}$ signify the scanning start and end frequencies, respectively, $T_{\text{swp}}$ denotes the scanning period and $\theta_J$ represents the initial phase.

{\bf\em 3) Frequency modulation (FM) jamming:}
it is expressed as:
\begin{equation}\label{eq:5}
    j(t) = \sum_{k=1}^{n} \sqrt{P_{J_k}} e^{j(2\pi f_{J_k} t + \beta_k\sin({2\pi f_{J_k} t}))}.
\end{equation}
FM interference signals also include both single-tone and multi-tone scenarios, with their carrier frequency influenced by the modulation factor $\beta_k$ and changing over time. Here, $P_{J_k}$ represents the power of the $k^{th}$ interference component, while $f_{J_k}$ denotes its frequency.

{\bf\em 4) Pulse jamming or distance measurement equipment (DME)-like jamming:}
it is expressed as:
\begin{equation}\label{eq:6}
    j(t) = \sqrt{P_J}p_\tau(t) \otimes \sum_{k=1}^{n} \delta(t-\frac{k}{f_{r_k}})e^{j(2\pi f_{J_k} t)}.
\end{equation}
%Pulse interference refers to interference signals that are active only within a specific time interval, and the ratio of active time to the total period is referred  as the duty cycle.
Pulse interference refers to signals that are active only within a specific time interval, with the ratio of active time to the total period termed the duty cycle.
Here, $P_J$ denotes the interference power, $p_\tau(t)$ denotes a rectangular pulse with a duty cycle of  $\tau$, $f_{r_k}$ is the repetition frequency of the pulse and 
$f_{J_k}$ is the interference frequency.

{\bf\em 5) Narrow band (NB) jamming:}
it is expressed as:
\begin{equation}\label{eq:7}
    j(t) = \sqrt{P_J}\cos{(2\pi f_J t + \beta \int_{0}^t n(\zeta) d\zeta + \theta_J)}.
\end{equation}
Narrow-band interference operates within a relatively narrow frequency range of the signal.
Here, $P_J$  represents the interference power, $\beta$ is the modulation index, and $n(\zeta)$ is a stationary random process with a mean of $0$ and a variance of $\sigma_{\zeta}^2$.

Following \cite{morales2019jammer}, this paper excludes wide band (WB) jamming from consideration due to the difficulty in detecting its presence. 
The expressions for these five types of interference signals are detailed in \cite{morales2019jammer}.
By applying short-time frequency spectrum transformation, the spectrogram of $r(t)$ is generated and rendered as a black-and-white image.

In this paper, we employ a FL strategy, whereby the central cloud server aggregates interference detector models from the $M = 5$ regions.
Ultimately, the interference detectors across regions share their model parameters with the central server, thus completing the recognition of interference signals.

\section{Federated Learning Methodology}
In conventional machine learning, centralized learning is a common methodology that focuses on training models on a central server.
FL is distinguished by its emphasis on privacy, similar to distributed learning, which spreads data across various nodes.
Many clients independently train model parameters and then collaborate with a central server to update the global model parameters.
FedAvg, a widely used FL algorithm, leverages this approach by allowing each client to independently train its local model. The global model parameters are then improved by calculating the weighted average of the local model parameters\cite{mcmahan2017communication}.

As demonstrated in \cite{wu2023jammer}, FL has shown promising results in the field of image classification, with its classification outcomes comparable to state-of-the-art centralized classification.
$M$ clients collaboratively train a neural classification network, the model is defined as follows: 
\begin{equation}\label{eq:8}
    \mathbf{y=h(X;\boldsymbol{\omega})},
\end{equation}
where $\mathbf{y} \in \mathbb{R}^C$ is the classification result, which consists of elements $p(y=l|\mathbf{X})$, $l \in \{AM,Chirp,FM,DME,NB,No\}$, $C$ is the number of interference types, $\mathbf{X}$ is the input image and $\boldsymbol{\omega}$ represents the parameters of the global model.
Each client contributes its local dataset $\mathcal{D}_m,m \in \{1,...,M\}$ to the overall dataset $\mathcal{D}$.
We denote the size of each dataset $\mathcal{D}_m$ as $D_m$ and the size of overall dataset $\mathcal{D}$ is $D=\sum_{m=1}^MD_m$.
Based on the FedAvg algorithm, the purpose of training is to minimize the loss function, which is defined as: 
\begin{equation}\label{eq:9}
    \min\limits_{\boldsymbol{\omega}} \mathcal{L}(\boldsymbol{\omega}) \quad \text{where} \quad \mathcal{L}(\boldsymbol{\omega}) = \sum_{m=1}^M \mathcal{F}_m(\boldsymbol{\omega})=\sum_{m=1}^M\frac{D_m}{D}\mathcal{L}_m(\boldsymbol{\omega}).
\end{equation}
Specifically, $\mathcal{L}(\boldsymbol{\omega})$ is the global model loss function, and $\mathcal{F}_m(\boldsymbol{\omega})$ is the local model loss function for zone $m$.
$\mathcal{L}_m(\boldsymbol{\omega})=\frac{1}{D_m}\sum_{n\in{\mathcal{D}_m} }f_n(\boldsymbol{\omega})$, in which $f_n(\boldsymbol{\omega})$ is the loss function of the sample $n$.
The optimization process of the Eq.\eqref{eq:9} can be expressed as follows.
Before the start of the first iteration, the central server initializes a global model.
During each iteration, the central server assigns the global model to each participating client in federated learning as its local model.
Subsequently, each client conducts training on its local model using its local dataset to update local parameters, as denoted in:
\begin{equation}\label{eq:10}
    \boldsymbol{\omega}_m^{t+1}=\arg \min_{\boldsymbol{\omega}}\mathcal{L}_{m,t}(\boldsymbol{\omega}).
\end{equation}
Following this, the central server aggregates the local model parameters by applying weight coefficients as:
\begin{equation}\label{eq:11}
    \boldsymbol{\omega}^{t+1}=\sum_{m=1}^{M} \frac{D_m}{D}\boldsymbol{\omega}_m^{t+1},
\end{equation}
where $\boldsymbol{\omega}_m^{t+1}$ is the updated local parameters and $\boldsymbol{\omega}^{t+1}$ is the updated global parameters. After the aggregation, the local model parameters together form the update for the global model.
The central server utilizes these updates to refresh the global model.
Ultimately, this iterative process is repeated a predefined number of times, indicated by $T$.

\section{Simulation Results}
\subsection{Preprocessing procedure}
In our experiments, we use the dataset from \cite{morales2019jammer}.
The dataset contains $61,800$ binary images with a resolution of $512 \times 512$ pixels and $600$ DPI (Dots Per Inch).
The dataset's images are spectrum plots of various interference signals, obtained via short-time Fourier transform.
The authors of \cite{morales2019jammer} used $6,000$ images, $1,000$ per jammer type for training, $1,800$ images, $300$ per type for validation, and $54,000$ images, $9,000$ images per type for testing.
To simplify training and save computational resources, we exclude the validation set in this study.
Consequently, The dataset consists exclusively of $75\%$ training and $25\%$ testing sets.
The training set comprises $10,800$ image, $1,800$ images per jammer type, while the testing set comprises $3600$ images, $600$ images per jammer type.
For faster training during data preprocessing, we downscale image dimensions from $512 \times 512$ to both $256 \times 256$ and $224 \times 224$.
\subsection{Data distribution}
\begin{figure}[h]
\centering
\includegraphics[width=0.5\textwidth]{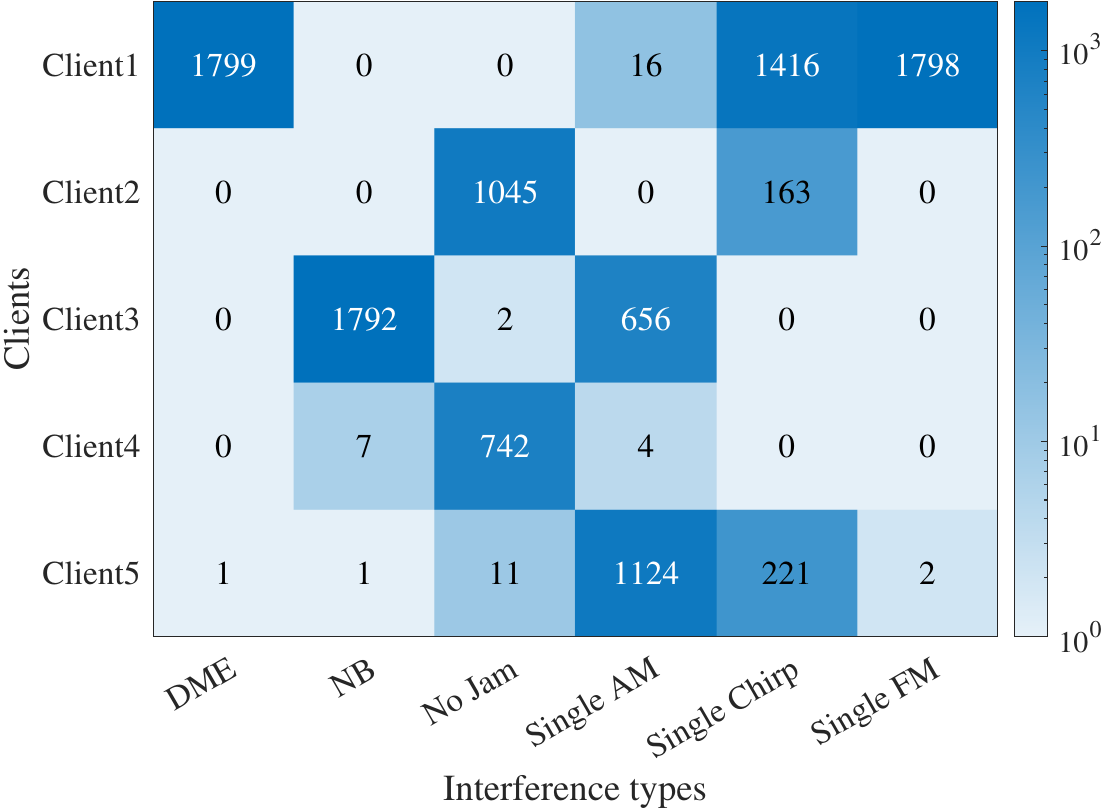}
\caption{Number of data points per class for each of the clients in non-IID scenario}
\label{fig:2}
\end{figure}

In this study, we model the diverse interference strategies of the movable UAV jammer on GNSS receivers across $M=5$ regions by intentionally creating a non-IID training dataset.
This approach ensures that movable UAV jammer emits specific interference patterns in each region, aligning with different interference strategies and fitting the system model.
Making the dataset non-IID effectively simulates the diversity and complexity of data from various real-world participants, enhancing the model's robustness and adaptability.
As discussed in \cite{wu2023jammer}, the authors achieved non-IID by modeling client data to conform to a Dirichlet distribution.
Each client receives different categories of interference signals based on probabilities generated by the Dirichlet distribution.
These probabilities are determined by the concentration parameter $\beta$, where a larger value results in a more uniform distribution with less disparity among components. 
Conversely, a smaller concentration parameter leads to greater disparities among components, causing the distribution to focus on specific components.
In this research, we select a concentration parameter of $0.1$ to highlight the data distribution imbalance among clients, ensuring that each client has only specific categories of interference signals.
Fig. \ref{fig:2} illustrates the distribution of data quantities corresponding to each interference category for each client when $M=5$.
Clearly, the data distribution shows significant diversity, with each client having only a limited number of interference categories. This accurately reflects the diverse interference strategies employed by drones in different regions.

Furthermore, this study examines the model's performance during training with client data sampled from an IID distribution, contrasting these results with non-IID data scenarios.
This comparative analysis enhances our understanding of the system model's resilience to interference, especially when client data distribution shows significant diversity.
Under IID conditions, where each of the client encounters an equal number of interference signals, the sample size totals $360$.

\subsection{Model Setting}
% 2-3 models to compare efficiency and accurancy
Authors in \cite{morales2019jammer} used a basic CNN to conduct classification tasks, yielding favorable outcomes.
In our study, we first adopt a similar CNN architecture, comprising a convolutional layer, a ReLU layer, a pooling layer, and a fully connected layer.
Specifically, the convolutional layer uses $16$ filters, each with dimensions of $12\times12\times1$.
The ReLU layer reduces redundant computations by keeping positive inputs and zeroing out negative ones.
The pooling layer has a size of $2\times2$. Subsequently, the fully connected layer integrates features across the network.
The softmax layer outputs class probabilities, enabling predictions across different categories. During training, we set a learning rate of $0.01$ and use the stochastic gradient descent (SGD) optimizer \cite{qian1999momentum}.

Furthermore, we utilize TL with the VGGNet architecture\cite{simonyan2014very} for classification. 
The VGG-16 model includes $13$ convolutional layers with $3\times3$ kernels, $5$ max-pooling layers, and $3$ fully connected layers.
In this study, we apply TL by adjusting the sixth fully connected layer's output size of VGG-16 to match our six classification categories.
We use both pre-trained and untrained network models for interference classification.
During training, we employed the Adam optimizer \cite{kingma2014adam} with a learning rate of $1\times 10^{-5}$, while using the cross-entropy function to calculate the loss.

\subsection{Results}
\begin{figure}[h]
\centering
\includegraphics[width=0.5\textwidth]{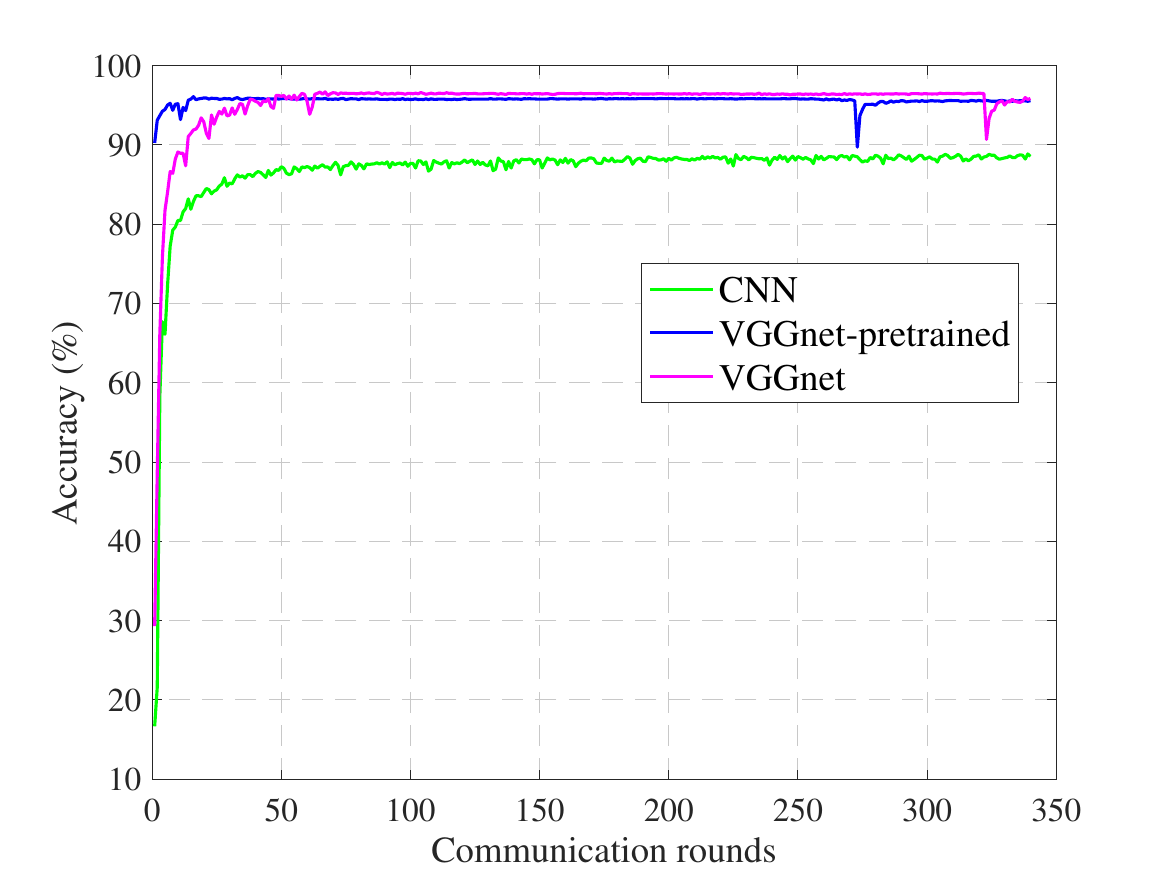}
\caption{Accuracy in $340$ rounds under different models}
\label{fig:3}
\end{figure}

% number of data points per classes for each client in non-iid scenario Fig.3 in reference
% Accurancy on different epochs
% C/N0 on different epochs (compare solo/iid FL/non-iid FL/centralized)
% different model accuracy comparison
Figure \ref{fig:3} shows a comparison of interference classification accuracy using three different network models.
Among these, the CNN model \cite{morales2019jammer} achieves an accuracy of $88.86\%$.
In contrast, the untrained VGGnet model improves accuracy by about $8\%$ to $96.69\%$.
Additionally, when comparing the convergence of pre-trained and untrained VGGnet models, the pre-trained model converge faster, achieving $95.92\%$ accuracy.
This demonstrates the quicker convergence of pre-trained models on target tasks in transfer learning.
Moreover, the untrained model's slightly higher convergence accuracy than the pre-trained model suggests that untrained models can adapt more flexibly to specific tasks.

\begin{figure}[h]
\centering
\includegraphics[width=0.5\textwidth]{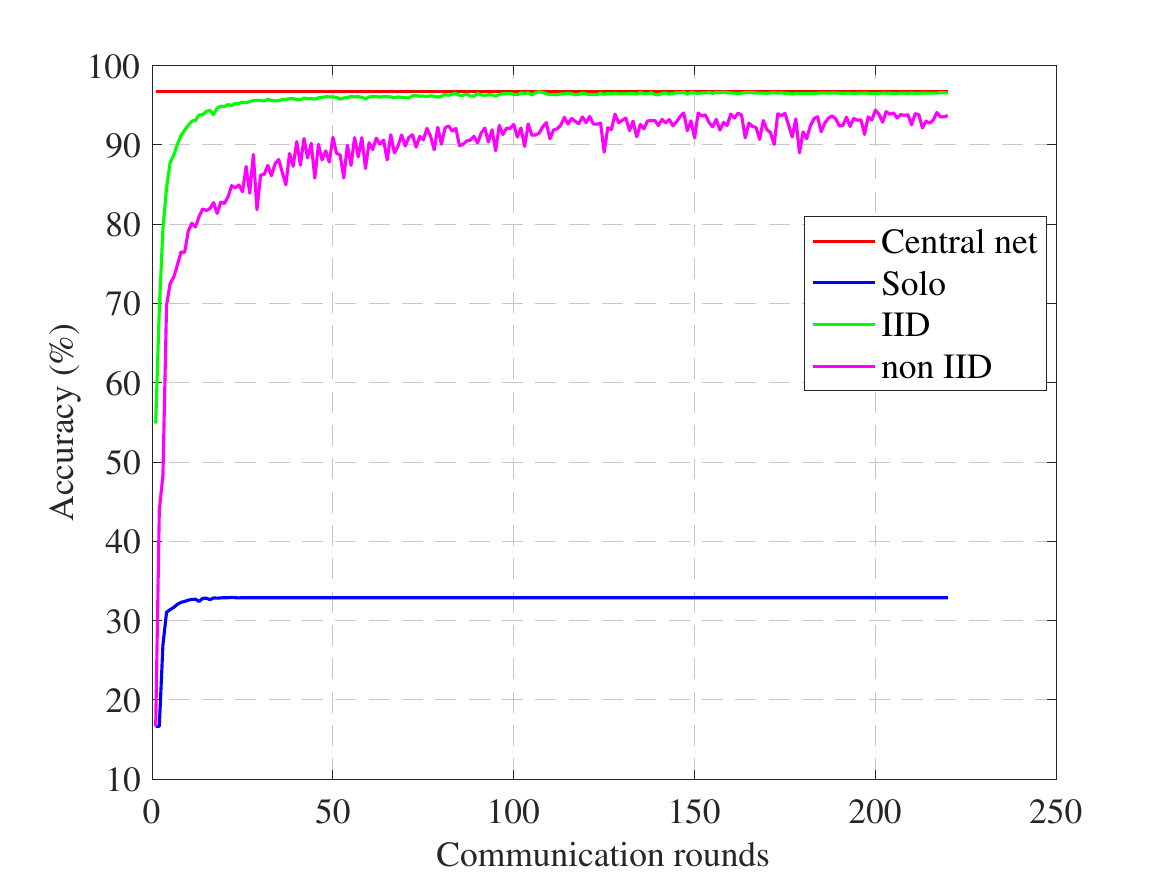}
\caption{FedAvg accuracy  under different data settings and Solo accuracy in $220$ rounds}
\label{fig:4}
\end{figure}

Figure \ref{fig:4} shows the classification accuracy of the FedAvg algorithm for two different data distributions and the accuracy in the Solo scenario.
In the Solo scenario, training occurs on data from a single region, with the dataset containing just two interference types and no model parameter exchange with other regions.
The accuracy of the centralized training model network serves as the baseline (with an accuracy of $96.69\%$). 
Both the FedAvg and Solo scenario use an untrained VGGnet model. Under IID conditions, the FedAvg algorithm's classification accuracy closely matches centralized training at about $96.69\%$.
However, with non-IID datasets, accuracy drops slightly to $94.38\%$, indicating the increased challenge of learning from diverse data distributions.
Additionally, in the Solo scenario, classification accuracy falls to $32.89\%$, only identifying the two types of interference in its dataset.
This shows that the FedAvg algorithm can recognize different interference types without direct data exchange, even as the UAV jammer's interference strategies change, unlike in the Solo scenario, where the classifier struggles with unencountered interference types.

\begin{figure}[h]
\centering
\includegraphics[width=0.5\textwidth]{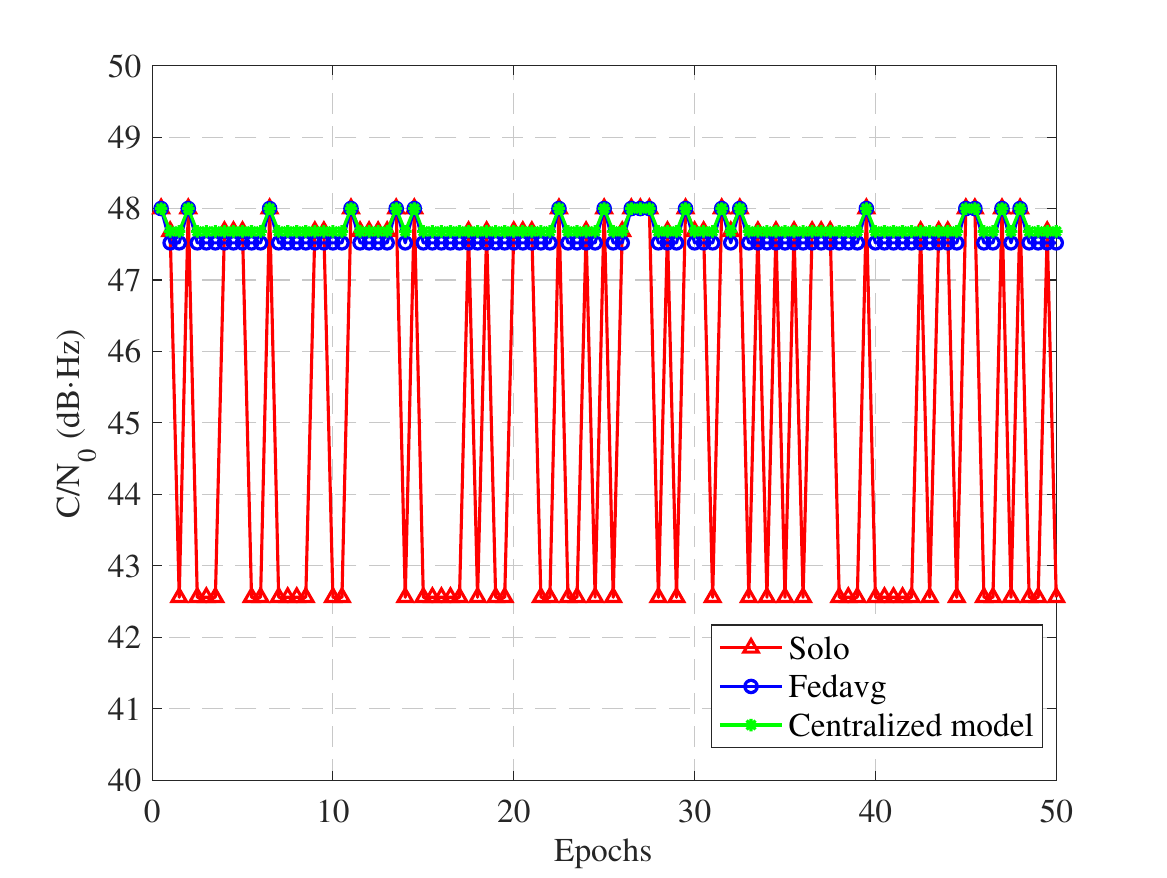}
\caption{Sum of $C/N_0$ for each zone under different classification models. }
\label{fig:5}
\end{figure}

Suppose that a movable UAV interference source in a given zone $m$ emits random types of interference at each moment.
Fig. \ref{fig:5} illustrates the $C/N_0$ values across various interference models and time instances.
Assume $C/{N_0}_{m,j}$ is $40\text{dB}\cdot \text{Hz}$ and $\tilde{C/{N_0}_{m,j}}$ is $48\text{dB}\cdot \text{Hz}$. Considering the non-uniform distribution of interference across regions, we assess the FedAvg model's performance with non-IID datasets.
With interferences, the $C/{N_0}_{m}$ value with the FedAvg algorithm is $47.52\text{dB}\cdot \text{Hz}$, comparable to centralized learning models at $47.68\text{dB}\cdot \text{Hz}$.
However, in the Solo scenario, facing three untrained interference types, the $C/{N_0}_{m}$ drops to $42.56\text{dB}\cdot \text{Hz}$, as expected.
This shows that the FedAvg algorithm improves the regional $C/N_0$ by about $5\text{dB}\cdot \text{Hz}$ over the Solo scenario in interference conditions.

In conclusion, our findings indicate that using the VGGnet model via TL improves interference classification accuracy by approximately $8\%$ over the convolutional network described in \cite{morales2019jammer}. Furthermore, when client datasets are IID, the FedAvg algorithm's classification accuracy closely matches that of centralized learning, with only a slight $3\%$ difference under the Dirichlet distribution.
Additionally, using the FedAvg algorithm in interference-affected regions can increase the regional $C/N_0$ by about $5\text{dB}\cdot \text{Hz}$ over the Solo scenario.
This approach not only protects regional users' privacy but also lessens the negative effects of interference on communication performance.

\section{Conclusion}
This paper presented a study on identifying and classifying GNSS interference signals from the movable UAV jammer in five wireless communication regions.
The simulation offered insights into classifying six potential interference spectrogram in GNSS systems under various models.
Using TL with VGGnet, we achieved an approximate $8\%$ improvement in classification accuracy over the CNN model\cite{morales2019jammer}.
Notably, pretrained networks are demonstrated to converge faster than untrained models.
Furthermore, the FL framework shows performance comparable to centralized learning, especially when datasets follow an IID pattern.
Additionally, the FedAvg approach, compared to the Solo scenario in individual regions, not only protects regional privacy but also effectively tackles interference recognition amidst evolving interference strategies, as this approach significantly enhanced the regional communication performance metric, $C/N_0$, by about $5\text{dB}\cdot \text{Hz}$.
% \section*{Acknowledgement}

\bibliographystyle{IEEEtran}
\bibliography{Reference}
\end{document}